\documentclass[aps,prl,reprint,superscriptaddress,showpacs]{revtex4-1}
\usepackage{graphicx,epstopdf}

\begin{document}

\title{Electrostatic Coupling between Two Surfaces of a Topological Insulator Nanodevice}

\author{Valla Fatemi}
\email[]{vfatemi@mit.edu}
\affiliation{Department of Physics, Massachusetts Institute of Technology, Cambridge, Massachusetts 02139, USA}

\author{Benjamin Hunt}
\affiliation{Department of Physics, Massachusetts Institute of Technology, Cambridge, Massachusetts 02139, USA}
\author{Hadar Steinberg}
\affiliation{Racah Institute of Physics, The Hebrew University, Jerusalem 91904, Israel}
\affiliation{Department of Physics, Massachusetts Institute of Technology, Cambridge, Massachusetts 02139, USA}
\author{Stephen L. Eltinge}
\affiliation{Department of Physics, Massachusetts Institute of Technology, Cambridge, Massachusetts 02139, USA}

\author{Fahad Mahmood}
\affiliation{Department of Physics, Massachusetts Institute of Technology, Cambridge, Massachusetts 02139, USA}
\author{Nicholas P. Butch}
\affiliation{Center for Neutron Research, National Institute of Standards and Technology, 100 Bureau Drive, MS 6100 Gaithersburg, MD 20899, USA}
\affiliation{Center for Nanophysics and Advanced Materials, Department of Physics, University of Maryland, College Park, MD 20742}
\affiliation{Lawrence Livermore National Laboratory 7000 East Avenue, Livermore, CA 94550, USA}
\author{Kenji Watanabe}
\affiliation{Advanced Materials Laboratory, National Institute for Materials Science, 1-1 Namiki, Tsukuba 305-0044, Japan}
\author{Takashi Taniguchi}
\affiliation{Advanced Materials Laboratory, National Institute for Materials Science, 1-1 Namiki, Tsukuba 305-0044, Japan}
\author{Nuh Gedik}
\affiliation{Department of Physics, Massachusetts Institute of Technology, Cambridge, Massachusetts 02139, USA}
\author{Ray Ashoori}
\affiliation{Department of Physics, Massachusetts Institute of Technology, Cambridge, Massachusetts 02139, USA}
\author{Pablo Jarillo-Herrero}
\affiliation{Department of Physics, Massachusetts Institute of Technology, Cambridge, Massachusetts 02139, USA}

\date{\today}

\begin{abstract}

We report on electronic transport measurements of dual-gated nano-devices
of the low-carrier density topological insulator $\mbox{Bi}_{1.5}\mbox{Sb}_{0.5}\mbox{Te}_{1.7}\mbox{Se}_{1.3}$.
In all devices the upper and lower surface states are independently
tunable to the Dirac point by the top and bottom gate electrodes.
In thin devices, electric fields are found to penetrate through the
bulk, indicating finite capacitive coupling between the surface states.
A charging model allows us to use the penetrating electric field as
a measurement of the inter-surface capacitance $C_{TI}$ and the surface
state energy-density relationship $\mu(n)$, which is found to be
consistent with independent ARPES measurements. At high magnetic fields,
increased field penetration through the surface states is observed,
strongly suggestive of the opening of a surface state band gap due
to broken time-reversal symmetry.

\end{abstract}


\maketitle

Three dimensional topological insulators (3D TIs) have been undergoing
intense theoretical and experimental research on the properties of
their unique surface states \cite{hasan2010colloquium,qi2011topological}.
The presence of bulk carriers has hampered experimental progress,
so a variety of crystal growth \cite{hor2009ptypebi2se3,zhang2011bandstructure,ren2010largebulk,ren2011optimizing,kong2011ambipolar,gehring2012growthof}
and in-situ charge displacement techniques \cite{steinberg2010surface,checkelsky2011bulkband,yuan2011liquidgated,kim2012surface,segawa2012ambipolar}
have been applied to suppress bulk conductivity. For example, quaternary
TI materials of the form $\mbox{Bi}_{2-x}\mbox{Sb}_{x}\mbox{Te}_{3-y}\mbox{Se}_{y}$
have a significantly suppressed bulk contribution to transport, reaching
large bulk resistivities and insulating-like temperature dependence
\cite{ren2011optimizing,arakane2012tunable,taskin2011observation}.
Furthermore, exfoliation or growth of thin crystals has been used
to achieve surface-dominated transport \cite{lee2012gatetuned,kong2011ambipolar,kim2012surface,xia2013indications,gehring2012growthof}.
However, amid the extensive experimental effort on TI device transport,
there is no study reporting independent control over the density of
both the upper and lower surface states in a single TI device. A full
understanding of transport phenomena in TIs, such as the quantum Hall
\cite{tilahun2011quantum,vafek2011quantum} and Josephon effects \cite{williams2012unconventional,veldhorst2012josephson,rokhinson2012thefractional},
will require independent tuning of the density of each surface state.
Additionally, proposals for topological exciton condensates explicitly
require fine tuning the density of both surfaces \cite{seradjeh2009exciton},
and finite displacement fields from two gates can affect the quantum
anomalous Hall effect in TI-based systems \cite{yu2010quantized,chang2013experimental}. 

In this Letter, we report electronic transport measurements of exfoliated
$\mbox{Bi}_{1.5}\mbox{Sb}_{0.5}\mbox{Te}_{1.7}\mbox{Se}_{1.3}$ (BSTS)
nanodevices with top and bottom gate electrodes. We show for the first
time that the chemical potential of the upper and lower surface states
can be controlled independently, resulting in different resistance
peaks when either surface chemical potential crosses the Dirac point.
For thin devices, we find signatures of finite capacitive coupling
between the surface states, consistent with fully depleted bulk states.
We explain the data through a charging model which incorporates the
finite density of states of the surface bands. Using angle-resolved
photoemission spectroscopy (ARPES) as a control measurement of the
surface state, this model allows us to measure the chemical potential
$\mu$ and charge density $n$ of a topological surface state as well
as the inter-surface capacitance $C_{TI}$. At high magnetic fields,
increased field penetration through the surface states is observed,
strongly suggestive of the opening of a surface state band gap. 

\begin{figure}[t]
\includegraphics{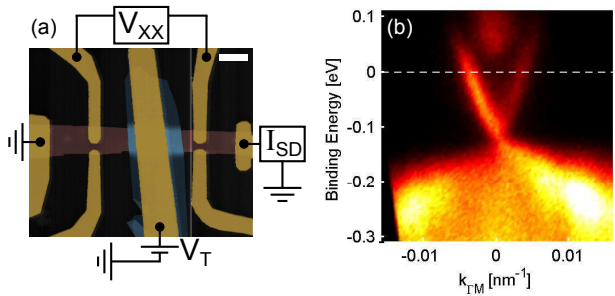}

\caption{\label{fig:1} (Color online) (a) Colorized AFM image of device A, including schematic
circuit elements describing the transport measurement. Red is BSTS,
blue is h-BN, and gold is Ti/Au (contacts and gate electrode). The
scale bar is 2 microns. (b) TrARPES measurement of a BSTS crystal.
The white line indicates the chemical potential.}
\end{figure}

BSTS was prepared by melting high purity samples of the constituent
elements in a sealed quartz ampoule under inert atmosphere. Sample
structure was confirmed by x-ray powder diffraction, and large single
crystals showed similar bulk transport behavior to previous reports
\cite{ren2011optimizing}. Static ARPES shows that the chemical potential
is inside the bulk band gap and that the Dirac point energy is above
the bulk valence band edge (see SM \cite{SMO}). Pump-probe time-resolved
ARPES (TrARPES) allows access to unoccupied states as shown in Fig.
\ref{fig:1}b \cite{sobota2012ultrafast,wang2012measurement}. The
Fermi velocity near the Dirac point is $v_{F}\approx3.2\times10^{5}\mbox{m/s}$,
and the band gap at room temperature is $E_{g}\approx240\mbox{ meV}$.
Note that the surface state dispersion is strongly electron-hole asymmetric.
These data are consistent with previous experiments \cite{arakane2012tunable,kim2014robustprotection}.

Thin flakes for transport studies were obtained by mechanical exfoliation
onto a doped silicon wafer with a 285nm thick thermal $\mbox{SiO}_{2}$
surface layer that serves as the bottom gate electrode and dielectric,
respectively. A thin layer of hexagonal boron nitride (h-BN) was mechanically
transferred on top to serve as the top gate dielectric \cite{dean2010boronnitride}.
Thermally evaporated Ti/Au layers were used to make ohmic contacts
and top gate electrodes. Atomic Force Microscopy was used to determine
the thickness of the BSTS and h-BN layers. For all data presented
here, a four-probe voltage measurement was used determine the 2D resistivity.
Here we report results measured on BSTS devices of different thicknesses:
device A is 42 nm, and device B is 82 nm. The behavior of device A
was reproduced in a third device \cite{SMO}. All three devices were
fabricated from flakes from the same exfoliation, and therefore from
the same region of the bulk crystal. Fig.  \ref{fig:1}a, shows an
AFM image of device A.

\begin{figure}[t]
\includegraphics{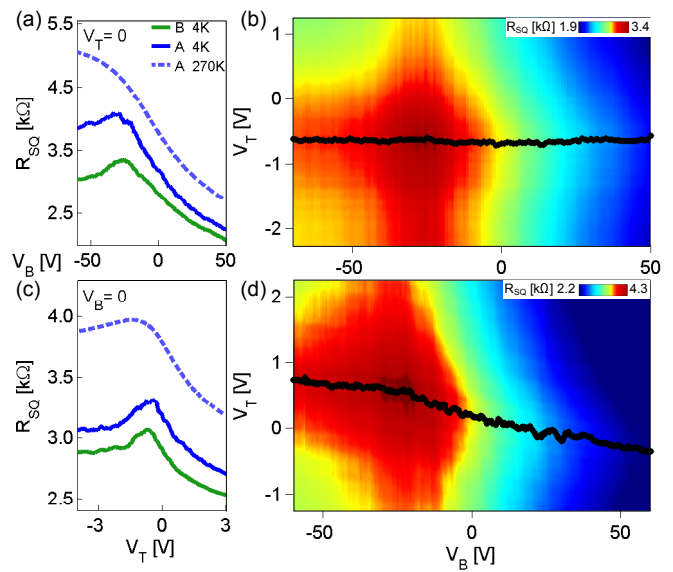}

\caption{\label{fig:2} (Color online) Gate-dependence of the resistivity of devices A and
B. (a) Bottom gate dependence of resistivity at $V_{T}=0$ at low
temperature (blue, green) and 270K (dashed) from cooldown 2. (c) Top
gate dependence of resistivity at $V_{B}=0$ at low temperature (blue,
green) and 270K (dashed) from cooldown 2. (b,d) 2D map of resistivity
while modulating both gate electrodes for devices B and A, respectively,
from cooldown 1. The black dots track the location of the upper surface
$\mbox{\ensuremath{R_{peak}}}$ at each $V_{B}$.}
\end{figure}

On devices A and B, both the top and bottom gates easily tune the
device through a resistance peak ($\mbox{\ensuremath{R_{peak}}}$)
by adjusting the applied voltages $V_{T}$ and $V_{B}$, respectively,
as shown in Fig. \ref{fig:2}a-b. $\mbox{\ensuremath{R_{peak}}}$
is associated with a minimum in carrier density (i.e. the surface
Dirac point), as confirmed via the Hall effect \cite{SMO}. Interestingly,
the top-gate $\mbox{\ensuremath{R_{peak}}}$ is observable up to room
temperature; in contrast, for the bottom gate $R(V_{B})$ changes
into a broad S-shape, consistent with gating studies of other TIs
using $\mbox{SiO}_{2}$ gate dielectrics \cite{checkelsky2011bulkband,xia2013indications,kong2011ambipolar}.
The disappearance of a distinct resistance peak in the limit of strong
disorder was predicted by recent theories for TI surface states with
electron-hole asymmetry \cite{adam2012twodimensional}, suggesting
that the difference in the field-effect behavior may be related to
the disorder profile at the interface. Strong differences in the disorder
profile at $\mbox{SiO}_{2}$ and h-BN interfaces have been observed
in graphene \cite{xue2011scanning}. 

Two-dimensional maps of the resistivity with respect to both top and
bottom gate voltage reveal a distinct difference in the behavior of
devices A and B, shown in Fig. \ref{fig:2}d and \ref{fig:2}c, respectively.
The black dots identify $V_{T}$, the top gate voltage at which $\mbox{\ensuremath{R_{peak}}}$
is found, at each $V_{B}$. We associate $V_{peak}$ with charge neutrality
of the upper surface state: $n_{U}=0$. For device B, $V_{peak}$
is independent of $V_{B}$, demonstrating no capacitive coupling between
the upper surface and the bottom gate electrode. The fact that thicker
devices do not have this capacitive coupling suggests that mobile
bulk electronic states exist in the interior. By contrast, $V_{peak}$
in device A is dependent on $V_{B}$. The observed relationship $V_{peak}(V_{B})$
means that there exists a finite and \emph{non-constant} capacitive
coupling between the upper surface and the bottom gate. This capacitive
coupling requires field penetration through the lower surface state
and the interior of the thinner crystal, which fail to completely
screen electric fields. The contrasting gating behavior of the devices
is corroborated by the temperature dependence of their resistivities
(see SM \cite{SMO}). We also note that while dual-gated TI devices
have been previously reported \cite{kim2012surface,checkelsky2012diracfermionmediated},
the devices reported here are unique in that the two surface states
are tuned independently and separately observed. 

\begin{figure}[t]
\includegraphics{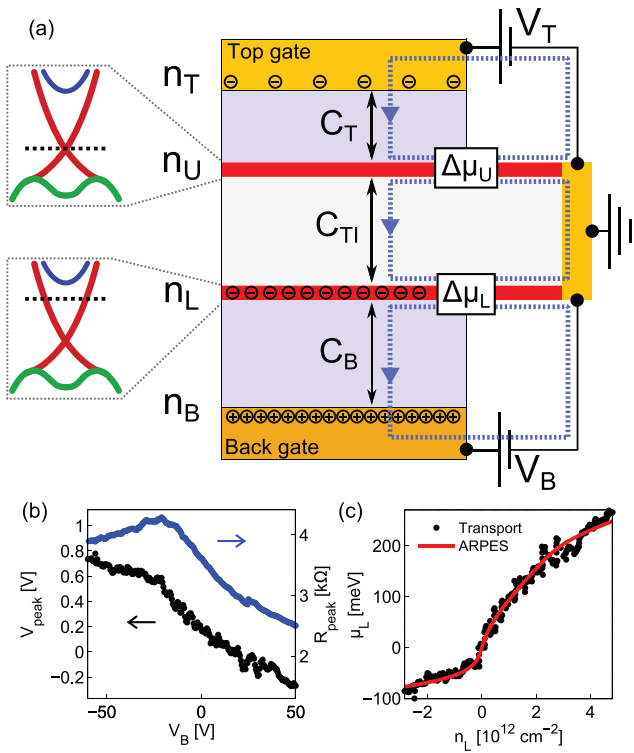}

\caption{\label{fig:3} (Color online) (a) Schematic of the charging model used in this study
with important parameters labeled. For comparison to the experiment,
the upper surface state is kept at charge neutrality while charge
is distributed between the lower surface state and the gate electrodes.
Three voltage loops indicated by the blue dashed lines are used in
deriving the charging model. (b) The position of the upper surface
$\mbox{\ensuremath{R_{peak}}}$ as a function of both gate voltages
$V_{peak}(V_{B})$ (left, black dots), and the resistivity at those
gate voltages $R_{peak}$ (right, blue), extracted from Fig. \ref{fig:2}d.
(c) The fit of the energy-density relationship as derived from ARPES
(red line) and from $V_{peak}(V_{B})$ (black dots).}
\end{figure}

Here we focus on the capacitive coupling between the bottom gate and
the upper surface in the thin crystal, and data regarding coupling
of the top gate and lower surface are presented in the SM \cite{SMO}.
The slope of $V_{peak}(V_{B})$ is a measure of the ratio of the capacitive
coupling of the bottom and top gates to the upper surface, which includes
partial screening of electric fields by the lower surface state. At
$V_{B}\sim-20\mbox{ V}$ the slope of $V_{peak}(V_{B})$ and the resistance
of the lower surface are simultaneously at a maximum, i.e. near the
Dirac point (see Fig. \ref{fig:3}b). This is consistent with a minimum
in the screening effectiveness of the lower surface state at the Dirac
point. Understanding this behavior quantitatively requires a detailed
charging model, which we discuss below. 

By considering the BSTS surface states as a grounded pair of 2D electronic
states, the general gating behavior can be understood via a charging
model construction originally developed for parallel graphene layers
\cite{kim2012directmeasurement}. This model is schematically represented
in Fig.  \ref{fig:3}a, where the important quantities are the applied
gate voltages ($V_{T}$, $V_{B}$), the geometric capacitances per
unit area of the gates ($C_{B}$, $C_{T}$), the inter-surface capacitance
per unit area ($C_{TI}$), the charge densities of the gate electrodes
($n_{T}$, $n_{B}$), and the charge density and chemical potentials
of the lower ($n_{L}$, $\mu_{L}$) and upper ($n_{U}$, $\mu_{U}$)
surface states. Four coupled equations completely describe the charging
of the system: one from charge neutrality, and three from Faraday's
law, which restricts the sum of voltage drops around a loop to equal
zero, which includes the change in chemical potential of the surface
states $\Delta\mu_{j}=\mu_{j}-\mu_{j}^{0}$, where $\mu_{j}^{0}$ is
the initial Fermi energy relative to the Dirac point for surface state
$j=U,L$ . A detailed derivation is provided in the SM \cite{SMO}.
For this study, we are interested in the condition that the chemical
potential at the upper surface is at the Dirac point. By setting $n_{U}=0$
and $\mu_{U}=0$, a useful pair of equations can be derived: 
\begin{eqnarray}
\mu_{L} & = & -\frac{C_{T}}{C_{TI}}eV_{T}^{\prime}\label{eq:muL}\\
\frac{1}{C_{B}}en_{L} & = & V_{B}^{\prime}+\left(\frac{1}{C_{B}}+\frac{1}{C_{TI}}\right)C_{T}V_{T}^{\prime},\label{eq:nL}
\end{eqnarray}
 where $V_{T,B}^{\prime}=V_{T,B}-V_{T,B}^{0}$, and $V_{T,B}^{0}$
are constants that depend on the initial densities and chemical potentials
of the two surfaces (see SM \cite{SMO}). Equations \ref{eq:muL}
and \ref{eq:nL} serve as a linear transformation from a trajectory
in gate voltage space (Fig. \ref{fig:3}b) to a relationship between
chemical potential and density for the lower surface state (Fig. \ref{fig:3}c). 

Experimentally, three unknowns remain: the inter-surface capacitance
$C_{TI}$ and the initial offset carrier densities of the upper and
lower surfaces $n_{L,U}^{0}$. To constrain these parameters, an independent
measurement of $\mu(n)$ is required. ARPES measurements of the surface
state band structure can be easily converted to a model for $E(n)$,
including an explicit treatment of the bulk states \cite{SMO}. A
three-parameter least-squares fit between the transformation of the
transport data and the ARPES model is performed and shown in Fig.
\ref{fig:3}c \cite{SMO}. The inter-layer capacitance from this fit
is $C_{TI}=740\pm20\mbox{ nF/c\ensuremath{m^{2}}}$, corresponding
to an effective bulk permittivity of $\kappa_{TI}\approx32$, comparable
to values for similar compounds \cite{collaboration:authorsantimony,petzelt1973farinfrared,richter1977araman}.
The initial electron densities of the upper and lower surface states
are found to be $n_{U}^{0}\approx-0.1\times10^{12}\mbox{c\ensuremath{m^{-2}}}$
and $n_{L}^{0}\approx1.2\times10^{12}\mbox{c\ensuremath{m^{-2}}}$,
which agrees well with values simply calculated from the magnitude
of $V_{T}$ and $V_{B}$ necessary to reach the resistance peaks. 

It is important to note that $C_{TI}$ can be affected in a few ways.
For example, localized electronic states could polarize, increasing
$C_{TI}$. As another possibility, low-density, poorly conducting
bulk states could weakly screen electric fields, reducing $C_{TI}$.
However, in the thin limit the surface states should efficiently screen
charged bulk impurities, resulting in an absence of charged puddles
of bulk states at charge neutrality for crystals of thickness $\lesssim70\mbox{ nm}$
\cite{skinner2013effects}. This length scale is consistent with the
observation that device B ($82\mbox{ nm}$ thick) appears to have
conducting states screening the two surfaces from each other.

\begin{figure}[t]
\includegraphics{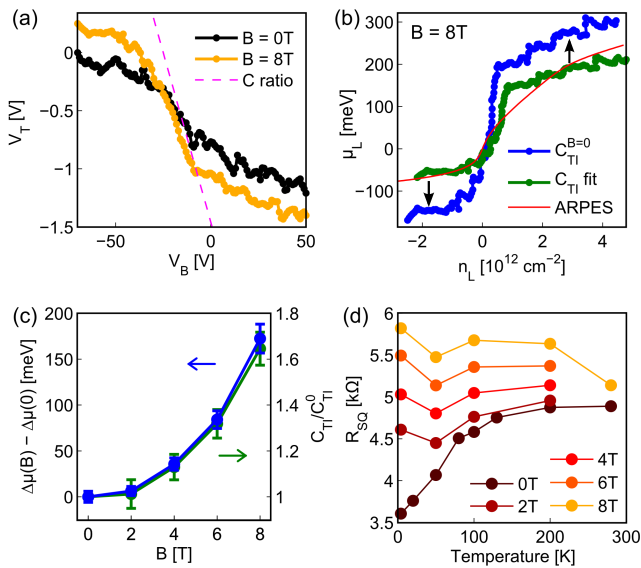}

\caption{\label{fig:4} Effect of high magnetic fields on the transport data.
(a) $V_{peak}(V_{B})$ at 0T and 8T from cooldown 2. For comparison,
the dashed pink line would be the gate-gate dependence if the lower
surface has no electronic states, given by a ratio of geometric capacitances:
$\mbox{C ratio}=-\frac{1}{C_{T}}\frac{C_{B}C_{TI}}{\left(C_{B}+C_{TI}\right)}$.
The transport data approaches this slope at 8T. (b) The extracted
energy-density relationship of the lower surface state at 8T for the
case of fixed inter-surface capacitance $C_{TI}=C_{TI}^{0}$ (blue)
and when using $C_{TI}$ as a fit parameter (green) to the zero-field
density of states (ARPES model, red curve). Arrows indicate increase in the total chemical potential change assuming fixed $C_{TI}$.  (c) The difference in
the total change of the chemical potential of the lower surface with
magnetic field (blue, left axis, error bars are the standard deviation
of possible values) and the best fit $C_{TI}$ as a function of magnetic
field (green, right axis, error bars are 90\% confidence intervals).
(d) The temperature dependence of the resistivity at different magnetic
fields and when both surfaces are at charge neutrality. }
\end{figure}

We now turn to the behavior of the thin device in high magnetic fields.
The Hall mobility of this sample is low, of order $200\mbox{ c\ensuremath{m^{2}}/(Vs)}$;
as a result, no evidence of Landau levels is found, and a clear $\mbox{\ensuremath{R_{peak}}}$
remains. Nevertheless, the charging behavior of the device changes
significantly at finite field. Fig.  \ref{fig:4}a shows $V_{T,peak}(V_{B})$
of the upper surface $\mbox{\ensuremath{R_{peak}}}$ at $B=0\mbox{ T}$
and $8\mbox{ T}$. $V_{peak}$ is affected by $V_{B}$ much more strongly
at 8T. Assuming $C_{TI}$ does not change, equations \ref{eq:muL}
and \ref{eq:nL} can be applied without changing parameters, as shown
in Fig.  \ref{fig:4}b (blue dots). For the same total change in charge
density, the total chemical potential change of the lower surface
is about 60\% larger. More precisely, the chemical potential appears
to change more rapidly at low carrier densities, indicating a distinctly
smaller thermodynamic density of states. Fig. \ref{fig:4}c (left
axis) shows the difference in total chemical potential change as a
function of magnetic field. The energy difference increases roughly
quadratically with magnetic field. A possible interpretation is that
the surface states develop a band gap that forms as a result of breaking
time-reversal symmetry. While a non-linear magnetic field dependence
would naively rule out a Zeeman-induced band gap, disorder will mask
this effect at low fields when the gap is small \cite{skinner2013effects},
causing a non-linear increase in the apparent gap in the density of
states. Detailed Shubnikov-de-Haas analysis of similar TI materials
estimate a surface $g$-factor in the range 40 to 80 \cite{taskin2011berryphase},
which would be too small to explain this effect, although the g-factor
has not yet been measured for this particular compound. 

We further observe that the temperature dependence of resistivity
also changes significantly at high magnetic fields. In Figure \ref{fig:4}d,
the temperature dependence of resistivity when both surfaces are at
charge neutrality changes from metallic-like at zero magnetic field
to non-metallic at high magnetic field, suggestive of a possible metal-insulator
transition. This is consistent with the formation of a gap in the
surface states with a high level of disorder. Similar non-metallic
resistivity vs temperature curves were observed in bilayer graphene
studies with similar band gaps in the high-disorder limit \cite{oostinga2008gateinduced}.

However, we cannot rule out the possibility of an inter-surface magneto-capacitance.
Restricting the model such that the total chemical potential change
is the same as at zero magnetic field (i.e. a field-independent \emph{average}
density of states, see green curve in Fig. \ref{fig:4}b), we find
that $C_{TI}$ must increase in magnetic field to compensate (Fig.
\ref{fig:4}c, right axis). $C_{TI}$ increases in a similar way as
the chemical potential difference because $\Delta\mu_{L}\cdot C_{TI}\propto\Delta V_{T}$,
as in equation \ref{eq:muL}. The raw bulk permittivity cannot explain
this change, because the optical phonon spectra of related TI compounds
show little change at similar magnetic fields \cite{sushkov2010farinfrared,laforge2010optical}.
Electronic contributions to $C_{TI}$ such as those mentioned earlier
(polarizable localized states or weakly screening bulk states) could
be modified by a magnetic field. In the supplement we show evidence
that the effects of temperature and magnetic field separately affect
$C_{TI}$ and $\mu_{L}(n)$, respectively \cite{SMO}, further suggesting
that the magnetic field is modifying the density of states and not
causing a magneto-capacitive effect. 

In summary, exfoliated nanoflakes of BSTS are of sufficiently low
total carrier density for both the upper and lower surface state densities
to be independently modulated by electrostatic gates and for electric
fields to penetrate through the bulk. Utilizing a model that captures
the charging of the system, we measure the inter-surface capacitance
$C_{TI}$ as well as the energy-density relationship $\mu(n)$ of
the surface states, which agrees well with independent ARPES measurements.
At high magnetic fields, increased field penetration is observed,
strongly suggestive of band gap opening in the lower surface state. 

This work was partly supported by the DOE, Basic Energy Sciences Office,
Division of Materials Sciences and Engineering, under award DE-SC0006418
(VF, SE, HS, and PJH), by the Gordon and Betty Moore Foundation grant GBMF2931 and the STC Center for Integrated Quantum Materials, NSF grant DMR-1231319 (BH and RCA), and by an MIT MRSEC Initiative under NSF award
DMR-0819762 (FM and NG). This work made use of the Materials Research
Science and Engineering Center Shared Experimental Facilities supported
by NSF under award DMR-0819762. Sample fabrication was performed partly
at the Harvard Center for Nanoscale Science supported by the NSF under
grant no. ECS-0335765. Sample synthesis and initial characterization
were performed under LDRD (Tracking Code 12-ERD-013) at Lawrence Livermore
National Laboratory (LLNL). LLNL is operated by Lawrence Livermore
National Security, LLC, for the US Department of Energy, National
Nuclear Security Administration, under Contract No. DE-AC52-07NA27344.
We thank A. Stern, Y. Baum, K. Burch, D. Drew, B. Skinner, A. Frenzel,
and J.D. Sanchez-Yamagishi for discussions and J. R. Jeffries for
performing x-ray diffraction measurements.

\bibliography{refs_BSTS_dualgate}

\begin{thebibliography}{42}%
\makeatletter
\providecommand \@ifxundefined [1]{%
 \@ifx{#1\undefined}
}%
\providecommand \@ifnum [1]{%
 \ifnum #1\expandafter \@firstoftwo
 \else \expandafter \@secondoftwo
 \fi
}%
\providecommand \@ifx [1]{%
 \ifx #1\expandafter \@firstoftwo
 \else \expandafter \@secondoftwo
 \fi
}%
\providecommand \natexlab [1]{#1}%
\providecommand \enquote  [1]{``#1''}%
\providecommand \bibnamefont  [1]{#1}%
\providecommand \bibfnamefont [1]{#1}%
\providecommand \citenamefont [1]{#1}%
\providecommand \href@noop [0]{\@secondoftwo}%
\providecommand \href [0]{\begingroup \@sanitize@url \@href}%
\providecommand \@href[1]{\@@startlink{#1}\@@href}%
\providecommand \@@href[1]{\endgroup#1\@@endlink}%
\providecommand \@sanitize@url [0]{\catcode `\\12\catcode `\$12\catcode
  `\&12\catcode `\#12\catcode `\^12\catcode `\_12\catcode `\%12\relax}%
\providecommand \@@startlink[1]{}%
\providecommand \@@endlink[0]{}%
\providecommand \url  [0]{\begingroup\@sanitize@url \@url }%
\providecommand \@url [1]{\endgroup\@href {#1}{\urlprefix }}%
\providecommand \urlprefix  [0]{URL }%
\providecommand \Eprint [0]{\href }%
\providecommand \doibase [0]{http://dx.doi.org/}%
\providecommand \selectlanguage [0]{\@gobble}%
\providecommand \bibinfo  [0]{\@secondoftwo}%
\providecommand \bibfield  [0]{\@secondoftwo}%
\providecommand \translation [1]{[#1]}%
\providecommand \BibitemOpen [0]{}%
\providecommand \bibitemStop [0]{}%
\providecommand \bibitemNoStop [0]{.\EOS\space}%
\providecommand \EOS [0]{\spacefactor3000\relax}%
\providecommand \BibitemShut  [1]{\csname bibitem#1\endcsname}%
\let\auto@bib@innerbib\@empty
%</preamble>
\bibitem [{\citenamefont {Hasan}\ and\ \citenamefont
  {Kane}(2010)}]{hasan2010colloquium}%
  \BibitemOpen
  \bibfield  {author} {\bibinfo {author} {\bibfnamefont {M.~Z.}\ \bibnamefont
  {Hasan}}\ and\ \bibinfo {author} {\bibfnamefont {C.~L.}\ \bibnamefont
  {Kane}},\ }\href {\doibase 10.1103/RevModPhys.82.3045} {\bibfield  {journal}
  {\bibinfo  {journal} {Rev. Mod. Phys.}\ }\textbf {\bibinfo {volume} {82}},\
  \bibinfo {pages} {3045} (\bibinfo {year} {2010})}\BibitemShut {NoStop}%
\bibitem [{\citenamefont {Qi}\ and\ \citenamefont
  {Zhang}(2011)}]{qi2011topological}%
  \BibitemOpen
  \bibfield  {author} {\bibinfo {author} {\bibfnamefont {X.-L.}\ \bibnamefont
  {Qi}}\ and\ \bibinfo {author} {\bibfnamefont {S.-C.}\ \bibnamefont {Zhang}},\
  }\href {\doibase 10.1103/RevModPhys.83.1057} {\bibfield  {journal} {\bibinfo
  {journal} {Rev. Mod. Phys.}\ }\textbf {\bibinfo {volume} {83}},\ \bibinfo
  {pages} {1057} (\bibinfo {year} {2011})}\BibitemShut {NoStop}%
\bibitem [{\citenamefont {Hor}\ \emph {et~al.}(2009)\citenamefont {Hor},
  \citenamefont {Richardella}, \citenamefont {Roushan}, \citenamefont {Xia},
  \citenamefont {Checkelsky}, \citenamefont {Yazdani}, \citenamefont {Hasan},
  \citenamefont {Ong},\ and\ \citenamefont {Cava}}]{hor2009ptypebi2se3}%
  \BibitemOpen
  \bibfield  {author} {\bibinfo {author} {\bibfnamefont {Y.~S.}\ \bibnamefont
  {Hor}}, \bibinfo {author} {\bibfnamefont {A.}~\bibnamefont {Richardella}},
  \bibinfo {author} {\bibfnamefont {P.}~\bibnamefont {Roushan}}, \bibinfo
  {author} {\bibfnamefont {Y.}~\bibnamefont {Xia}}, \bibinfo {author}
  {\bibfnamefont {J.~G.}\ \bibnamefont {Checkelsky}}, \bibinfo {author}
  {\bibfnamefont {A.}~\bibnamefont {Yazdani}}, \bibinfo {author} {\bibfnamefont
  {M.~Z.}\ \bibnamefont {Hasan}}, \bibinfo {author} {\bibfnamefont {N.~P.}\
  \bibnamefont {Ong}}, \ and\ \bibinfo {author} {\bibfnamefont {R.~J.}\
  \bibnamefont {Cava}},\ }\href {\doibase 10.1103/PhysRevB.79.195208}
  {\bibfield  {journal} {\bibinfo  {journal} {Phys. Rev. B}\ }\textbf {\bibinfo
  {volume} {79}},\ \bibinfo {pages} {195208} (\bibinfo {year}
  {2009})}\BibitemShut {NoStop}%
\bibitem [{\citenamefont {Zhang}\ \emph {et~al.}(2011)\citenamefont {Zhang},
  \citenamefont {Chang}, \citenamefont {Zhang}, \citenamefont {Wen},
  \citenamefont {Feng}, \citenamefont {Li}, \citenamefont {Liu}, \citenamefont
  {He}, \citenamefont {Wang}, \citenamefont {Chen}, \citenamefont {Xue},
  \citenamefont {Ma},\ and\ \citenamefont {Wang}}]{zhang2011bandstructure}%
  \BibitemOpen
  \bibfield  {author} {\bibinfo {author} {\bibfnamefont {J.}~\bibnamefont
  {Zhang}}, \bibinfo {author} {\bibfnamefont {C.-Z.}\ \bibnamefont {Chang}},
  \bibinfo {author} {\bibfnamefont {Z.}~\bibnamefont {Zhang}}, \bibinfo
  {author} {\bibfnamefont {J.}~\bibnamefont {Wen}}, \bibinfo {author}
  {\bibfnamefont {X.}~\bibnamefont {Feng}}, \bibinfo {author} {\bibfnamefont
  {K.}~\bibnamefont {Li}}, \bibinfo {author} {\bibfnamefont {M.}~\bibnamefont
  {Liu}}, \bibinfo {author} {\bibfnamefont {K.}~\bibnamefont {He}}, \bibinfo
  {author} {\bibfnamefont {L.}~\bibnamefont {Wang}}, \bibinfo {author}
  {\bibfnamefont {X.}~\bibnamefont {Chen}}, \bibinfo {author} {\bibfnamefont
  {Q.-K.}\ \bibnamefont {Xue}}, \bibinfo {author} {\bibfnamefont
  {X.}~\bibnamefont {Ma}}, \ and\ \bibinfo {author} {\bibfnamefont
  {Y.}~\bibnamefont {Wang}},\ }\href {\doibase 10.1038/ncomms1588} {\bibfield
  {journal} {\bibinfo  {journal} {Nat Commun}\ }\textbf {\bibinfo {volume}
  {2}},\ \bibinfo {pages} {574} (\bibinfo {year} {2011})}\BibitemShut {NoStop}%
\bibitem [{\citenamefont {Ren}\ \emph {et~al.}(2010)\citenamefont {Ren},
  \citenamefont {Taskin}, \citenamefont {Sasaki}, \citenamefont {Segawa},\ and\
  \citenamefont {Ando}}]{ren2010largebulk}%
  \BibitemOpen
  \bibfield  {author} {\bibinfo {author} {\bibfnamefont {Z.}~\bibnamefont
  {Ren}}, \bibinfo {author} {\bibfnamefont {A.~A.}\ \bibnamefont {Taskin}},
  \bibinfo {author} {\bibfnamefont {S.}~\bibnamefont {Sasaki}}, \bibinfo
  {author} {\bibfnamefont {K.}~\bibnamefont {Segawa}}, \ and\ \bibinfo {author}
  {\bibfnamefont {Y.}~\bibnamefont {Ando}},\ }\href {\doibase
  10.1103/PhysRevB.82.241306} {\bibfield  {journal} {\bibinfo  {journal} {Phys.
  Rev. B}\ }\textbf {\bibinfo {volume} {82}},\ \bibinfo {pages} {241306}
  (\bibinfo {year} {2010})}\BibitemShut {NoStop}%
\bibitem [{\citenamefont {Ren}\ \emph {et~al.}(2011)\citenamefont {Ren},
  \citenamefont {Taskin}, \citenamefont {Sasaki}, \citenamefont {Segawa},\ and\
  \citenamefont {Ando}}]{ren2011optimizing}%
  \BibitemOpen
  \bibfield  {author} {\bibinfo {author} {\bibfnamefont {Z.}~\bibnamefont
  {Ren}}, \bibinfo {author} {\bibfnamefont {A.~A.}\ \bibnamefont {Taskin}},
  \bibinfo {author} {\bibfnamefont {S.}~\bibnamefont {Sasaki}}, \bibinfo
  {author} {\bibfnamefont {K.}~\bibnamefont {Segawa}}, \ and\ \bibinfo {author}
  {\bibfnamefont {Y.}~\bibnamefont {Ando}},\ }\href {\doibase
  10.1103/PhysRevB.84.165311} {\bibfield  {journal} {\bibinfo  {journal} {Phys.
  Rev. B}\ }\textbf {\bibinfo {volume} {84}},\ \bibinfo {pages} {165311}
  (\bibinfo {year} {2011})}\BibitemShut {NoStop}%
\bibitem [{\citenamefont {Kong}\ \emph {et~al.}(2011)\citenamefont {Kong},
  \citenamefont {Chen}, \citenamefont {Cha}, \citenamefont {Zhang},
  \citenamefont {Analytis}, \citenamefont {Lai}, \citenamefont {Liu},
  \citenamefont {Hong}, \citenamefont {Koski}, \citenamefont {Mo},
  \citenamefont {Hussain}, \citenamefont {Fisher}, \citenamefont {Shen},\ and\
  \citenamefont {Cui}}]{kong2011ambipolar}%
  \BibitemOpen
  \bibfield  {author} {\bibinfo {author} {\bibfnamefont {D.}~\bibnamefont
  {Kong}}, \bibinfo {author} {\bibfnamefont {Y.}~\bibnamefont {Chen}}, \bibinfo
  {author} {\bibfnamefont {J.~J.}\ \bibnamefont {Cha}}, \bibinfo {author}
  {\bibfnamefont {Q.}~\bibnamefont {Zhang}}, \bibinfo {author} {\bibfnamefont
  {J.~G.}\ \bibnamefont {Analytis}}, \bibinfo {author} {\bibfnamefont
  {K.}~\bibnamefont {Lai}}, \bibinfo {author} {\bibfnamefont {Z.}~\bibnamefont
  {Liu}}, \bibinfo {author} {\bibfnamefont {S.~S.}\ \bibnamefont {Hong}},
  \bibinfo {author} {\bibfnamefont {K.~J.}\ \bibnamefont {Koski}}, \bibinfo
  {author} {\bibfnamefont {S.-K.}\ \bibnamefont {Mo}}, \bibinfo {author}
  {\bibfnamefont {Z.}~\bibnamefont {Hussain}}, \bibinfo {author} {\bibfnamefont
  {I.~R.}\ \bibnamefont {Fisher}}, \bibinfo {author} {\bibfnamefont {Z.-X.}\
  \bibnamefont {Shen}}, \ and\ \bibinfo {author} {\bibfnamefont
  {Y.}~\bibnamefont {Cui}},\ }\href {\doibase 10.1038/nnano.2011.172}
  {\bibfield  {journal} {\bibinfo  {journal} {Nat Nano}\ }\textbf {\bibinfo
  {volume} {6}},\ \bibinfo {pages} {705} (\bibinfo {year} {2011})}\BibitemShut
  {NoStop}%
\bibitem [{\citenamefont {Gehring}\ \emph {et~al.}(2012)\citenamefont
  {Gehring}, \citenamefont {Gao}, \citenamefont {Burghard},\ and\ \citenamefont
  {Kern}}]{gehring2012growthof}%
  \BibitemOpen
  \bibfield  {author} {\bibinfo {author} {\bibfnamefont {P.}~\bibnamefont
  {Gehring}}, \bibinfo {author} {\bibfnamefont {B.~F.}\ \bibnamefont {Gao}},
  \bibinfo {author} {\bibfnamefont {M.}~\bibnamefont {Burghard}}, \ and\
  \bibinfo {author} {\bibfnamefont {K.}~\bibnamefont {Kern}},\ }\href {\doibase
  10.1021/nl3019802} {\bibfield  {journal} {\bibinfo  {journal} {Nano Lett.}\
  }\textbf {\bibinfo {volume} {12}},\ \bibinfo {pages} {5137} (\bibinfo {year}
  {2012})}\BibitemShut {NoStop}%
\bibitem [{\citenamefont {Steinberg}\ \emph {et~al.}(2010)\citenamefont
  {Steinberg}, \citenamefont {Gardner}, \citenamefont {Lee},\ and\
  \citenamefont {Jarillo-Herrero}}]{steinberg2010surface}%
  \BibitemOpen
  \bibfield  {author} {\bibinfo {author} {\bibfnamefont {H.}~\bibnamefont
  {Steinberg}}, \bibinfo {author} {\bibfnamefont {D.~R.}\ \bibnamefont
  {Gardner}}, \bibinfo {author} {\bibfnamefont {Y.~S.}\ \bibnamefont {Lee}}, \
  and\ \bibinfo {author} {\bibfnamefont {P.}~\bibnamefont {Jarillo-Herrero}},\
  }\href {\doibase 10.1021/nl1032183} {\bibfield  {journal} {\bibinfo
  {journal} {Nano Lett.}\ }\textbf {\bibinfo {volume} {10}},\ \bibinfo {pages}
  {5032} (\bibinfo {year} {2010})}\BibitemShut {NoStop}%
\bibitem [{\citenamefont {Checkelsky}\ \emph {et~al.}(2011)\citenamefont
  {Checkelsky}, \citenamefont {Hor}, \citenamefont {Cava},\ and\ \citenamefont
  {Ong}}]{checkelsky2011bulkband}%
  \BibitemOpen
  \bibfield  {author} {\bibinfo {author} {\bibfnamefont {J.~G.}\ \bibnamefont
  {Checkelsky}}, \bibinfo {author} {\bibfnamefont {Y.~S.}\ \bibnamefont {Hor}},
  \bibinfo {author} {\bibfnamefont {R.~J.}\ \bibnamefont {Cava}}, \ and\
  \bibinfo {author} {\bibfnamefont {N.~P.}\ \bibnamefont {Ong}},\ }\href
  {\doibase 10.1103/PhysRevLett.106.196801} {\bibfield  {journal} {\bibinfo
  {journal} {Phys. Rev. Lett.}\ }\textbf {\bibinfo {volume} {106}},\ \bibinfo
  {pages} {196801} (\bibinfo {year} {2011})}\BibitemShut {NoStop}%
\bibitem [{\citenamefont {Yuan}\ \emph {et~al.}(2011)\citenamefont {Yuan},
  \citenamefont {Liu}, \citenamefont {Shimotani}, \citenamefont {Guo},
  \citenamefont {Chen}, \citenamefont {Xue},\ and\ \citenamefont
  {Iwasa}}]{yuan2011liquidgated}%
  \BibitemOpen
  \bibfield  {author} {\bibinfo {author} {\bibfnamefont {H.}~\bibnamefont
  {Yuan}}, \bibinfo {author} {\bibfnamefont {H.}~\bibnamefont {Liu}}, \bibinfo
  {author} {\bibfnamefont {H.}~\bibnamefont {Shimotani}}, \bibinfo {author}
  {\bibfnamefont {H.}~\bibnamefont {Guo}}, \bibinfo {author} {\bibfnamefont
  {M.}~\bibnamefont {Chen}}, \bibinfo {author} {\bibfnamefont {Q.}~\bibnamefont
  {Xue}}, \ and\ \bibinfo {author} {\bibfnamefont {Y.}~\bibnamefont {Iwasa}},\
  }\href {\doibase 10.1021/nl201561u} {\bibfield  {journal} {\bibinfo
  {journal} {Nano Lett.}\ }\textbf {\bibinfo {volume} {11}},\ \bibinfo {pages}
  {2601} (\bibinfo {year} {2011})}\BibitemShut {NoStop}%
\bibitem [{\citenamefont {Kim}\ \emph {et~al.}(2012{\natexlab{a}})\citenamefont
  {Kim}, \citenamefont {Cho}, \citenamefont {Butch}, \citenamefont {Syers},
  \citenamefont {Kirshenbaum}, \citenamefont {Adam}, \citenamefont {Paglione},\
  and\ \citenamefont {Fuhrer}}]{kim2012surface}%
  \BibitemOpen
  \bibfield  {author} {\bibinfo {author} {\bibfnamefont {D.}~\bibnamefont
  {Kim}}, \bibinfo {author} {\bibfnamefont {S.}~\bibnamefont {Cho}}, \bibinfo
  {author} {\bibfnamefont {N.~P.}\ \bibnamefont {Butch}}, \bibinfo {author}
  {\bibfnamefont {P.}~\bibnamefont {Syers}}, \bibinfo {author} {\bibfnamefont
  {K.}~\bibnamefont {Kirshenbaum}}, \bibinfo {author} {\bibfnamefont
  {S.}~\bibnamefont {Adam}}, \bibinfo {author} {\bibfnamefont {J.}~\bibnamefont
  {Paglione}}, \ and\ \bibinfo {author} {\bibfnamefont {M.~S.}\ \bibnamefont
  {Fuhrer}},\ }\href {\doibase 10.1038/nphys2286} {\bibfield  {journal}
  {\bibinfo  {journal} {Nat Phys}\ }\textbf {\bibinfo {volume} {8}},\ \bibinfo
  {pages} {459} (\bibinfo {year} {2012}{\natexlab{a}})}\BibitemShut {NoStop}%
\bibitem [{\citenamefont {Segawa}\ \emph {et~al.}(2012)\citenamefont {Segawa},
  \citenamefont {Ren}, \citenamefont {Sasaki}, \citenamefont {Tsuda},
  \citenamefont {Kuwabata},\ and\ \citenamefont {Ando}}]{segawa2012ambipolar}%
  \BibitemOpen
  \bibfield  {author} {\bibinfo {author} {\bibfnamefont {K.}~\bibnamefont
  {Segawa}}, \bibinfo {author} {\bibfnamefont {Z.}~\bibnamefont {Ren}},
  \bibinfo {author} {\bibfnamefont {S.}~\bibnamefont {Sasaki}}, \bibinfo
  {author} {\bibfnamefont {T.}~\bibnamefont {Tsuda}}, \bibinfo {author}
  {\bibfnamefont {S.}~\bibnamefont {Kuwabata}}, \ and\ \bibinfo {author}
  {\bibfnamefont {Y.}~\bibnamefont {Ando}},\ }\href {\doibase
  10.1103/PhysRevB.86.075306} {\bibfield  {journal} {\bibinfo  {journal} {Phys.
  Rev. B}\ }\textbf {\bibinfo {volume} {86}},\ \bibinfo {pages} {075306}
  (\bibinfo {year} {2012})}\BibitemShut {NoStop}%
\bibitem [{\citenamefont {Arakane}\ \emph {et~al.}(2012)\citenamefont
  {Arakane}, \citenamefont {Sato}, \citenamefont {Souma}, \citenamefont
  {Kosaka}, \citenamefont {Nakayama}, \citenamefont {Komatsu}, \citenamefont
  {Takahashi}, \citenamefont {Ren}, \citenamefont {Segawa},\ and\ \citenamefont
  {Ando}}]{arakane2012tunable}%
  \BibitemOpen
  \bibfield  {author} {\bibinfo {author} {\bibfnamefont {T.}~\bibnamefont
  {Arakane}}, \bibinfo {author} {\bibfnamefont {T.}~\bibnamefont {Sato}},
  \bibinfo {author} {\bibfnamefont {S.}~\bibnamefont {Souma}}, \bibinfo
  {author} {\bibfnamefont {K.}~\bibnamefont {Kosaka}}, \bibinfo {author}
  {\bibfnamefont {K.}~\bibnamefont {Nakayama}}, \bibinfo {author}
  {\bibfnamefont {M.}~\bibnamefont {Komatsu}}, \bibinfo {author} {\bibfnamefont
  {T.}~\bibnamefont {Takahashi}}, \bibinfo {author} {\bibfnamefont
  {Z.}~\bibnamefont {Ren}}, \bibinfo {author} {\bibfnamefont {K.}~\bibnamefont
  {Segawa}}, \ and\ \bibinfo {author} {\bibfnamefont {Y.}~\bibnamefont
  {Ando}},\ }\href {\doibase 10.1038/ncomms1639} {\bibfield  {journal}
  {\bibinfo  {journal} {Nat. Comm.}\ }\textbf {\bibinfo {volume} {3}},\
  \bibinfo {pages} {636} (\bibinfo {year} {2012})}\BibitemShut {NoStop}%
\bibitem [{\citenamefont {Taskin}\ \emph {et~al.}(2011)\citenamefont {Taskin},
  \citenamefont {Ren}, \citenamefont {Sasaki}, \citenamefont {Segawa},\ and\
  \citenamefont {Ando}}]{taskin2011observation}%
  \BibitemOpen
  \bibfield  {author} {\bibinfo {author} {\bibfnamefont {A.~A.}\ \bibnamefont
  {Taskin}}, \bibinfo {author} {\bibfnamefont {Z.}~\bibnamefont {Ren}},
  \bibinfo {author} {\bibfnamefont {S.}~\bibnamefont {Sasaki}}, \bibinfo
  {author} {\bibfnamefont {K.}~\bibnamefont {Segawa}}, \ and\ \bibinfo {author}
  {\bibfnamefont {Y.}~\bibnamefont {Ando}},\ }\href {\doibase
  10.1103/PhysRevLett.107.016801} {\bibfield  {journal} {\bibinfo  {journal}
  {Phys. Rev. Lett.}\ }\textbf {\bibinfo {volume} {107}} (\bibinfo {year}
  {2011}),\ 10.1103/PhysRevLett.107.016801}\BibitemShut {NoStop}%
\bibitem [{\citenamefont {Lee}\ \emph {et~al.}(2012)\citenamefont {Lee},
  \citenamefont {Park}, \citenamefont {Lee}, \citenamefont {Kim},\ and\
  \citenamefont {Lee}}]{lee2012gatetuned}%
  \BibitemOpen
  \bibfield  {author} {\bibinfo {author} {\bibfnamefont {J.}~\bibnamefont
  {Lee}}, \bibinfo {author} {\bibfnamefont {J.}~\bibnamefont {Park}}, \bibinfo
  {author} {\bibfnamefont {J.-H.}\ \bibnamefont {Lee}}, \bibinfo {author}
  {\bibfnamefont {J.~S.}\ \bibnamefont {Kim}}, \ and\ \bibinfo {author}
  {\bibfnamefont {H.-J.}\ \bibnamefont {Lee}},\ }\href {\doibase
  10.1103/PhysRevB.86.245321} {\bibfield  {journal} {\bibinfo  {journal} {Phys.
  Rev. B}\ }\textbf {\bibinfo {volume} {86}} (\bibinfo {year} {2012}),\
  10.1103/PhysRevB.86.245321}\BibitemShut {NoStop}%
\bibitem [{\citenamefont {Xia}\ \emph {et~al.}(2013)\citenamefont {Xia},
  \citenamefont {Ren}, \citenamefont {Sulaev}, \citenamefont {Liu},
  \citenamefont {Shen},\ and\ \citenamefont {Wang}}]{xia2013indications}%
  \BibitemOpen
  \bibfield  {author} {\bibinfo {author} {\bibfnamefont {B.}~\bibnamefont
  {Xia}}, \bibinfo {author} {\bibfnamefont {P.}~\bibnamefont {Ren}}, \bibinfo
  {author} {\bibfnamefont {A.}~\bibnamefont {Sulaev}}, \bibinfo {author}
  {\bibfnamefont {P.}~\bibnamefont {Liu}}, \bibinfo {author} {\bibfnamefont
  {S.-Q.}\ \bibnamefont {Shen}}, \ and\ \bibinfo {author} {\bibfnamefont
  {L.}~\bibnamefont {Wang}},\ }\href {\doibase 10.1103/PhysRevB.87.085442}
  {\bibfield  {journal} {\bibinfo  {journal} {Phys. Rev. B}\ }\textbf {\bibinfo
  {volume} {87}},\ \bibinfo {pages} {085442} (\bibinfo {year}
  {2013})}\BibitemShut {NoStop}%
\bibitem [{\citenamefont {Tilahun}\ \emph {et~al.}(2011)\citenamefont
  {Tilahun}, \citenamefont {Lee}, \citenamefont {Hankiewicz},\ and\
  \citenamefont {{MacDonald}}}]{tilahun2011quantum}%
  \BibitemOpen
  \bibfield  {author} {\bibinfo {author} {\bibfnamefont {D.}~\bibnamefont
  {Tilahun}}, \bibinfo {author} {\bibfnamefont {B.}~\bibnamefont {Lee}},
  \bibinfo {author} {\bibfnamefont {E.~M.}\ \bibnamefont {Hankiewicz}}, \ and\
  \bibinfo {author} {\bibfnamefont {A.~H.}\ \bibnamefont {{MacDonald}}},\
  }\href {\doibase 10.1103/PhysRevLett.107.246401} {\bibfield  {journal}
  {\bibinfo  {journal} {Phys. Rev. Lett.}\ }\textbf {\bibinfo {volume} {107}},\
  \bibinfo {pages} {246401} (\bibinfo {year} {2011})}\BibitemShut {NoStop}%
\bibitem [{\citenamefont {Vafek}(2011)}]{vafek2011quantum}%
  \BibitemOpen
  \bibfield  {author} {\bibinfo {author} {\bibfnamefont {O.}~\bibnamefont
  {Vafek}},\ }\href {\doibase 10.1103/PhysRevB.84.245417} {\bibfield  {journal}
  {\bibinfo  {journal} {Phys. Rev. B}\ }\textbf {\bibinfo {volume} {84}},\
  \bibinfo {pages} {245417} (\bibinfo {year} {2011})}\BibitemShut {NoStop}%
\bibitem [{\citenamefont {Williams}\ \emph {et~al.}(2012)\citenamefont
  {Williams}, \citenamefont {Bestwick}, \citenamefont {Gallagher},
  \citenamefont {Hong}, \citenamefont {Cui}, \citenamefont {Bleich},
  \citenamefont {Analytis}, \citenamefont {Fisher},\ and\ \citenamefont
  {Goldhaber-Gordon}}]{williams2012unconventional}%
  \BibitemOpen
  \bibfield  {author} {\bibinfo {author} {\bibfnamefont {J.~R.}\ \bibnamefont
  {Williams}}, \bibinfo {author} {\bibfnamefont {A.~J.}\ \bibnamefont
  {Bestwick}}, \bibinfo {author} {\bibfnamefont {P.}~\bibnamefont {Gallagher}},
  \bibinfo {author} {\bibfnamefont {S.~S.}\ \bibnamefont {Hong}}, \bibinfo
  {author} {\bibfnamefont {Y.}~\bibnamefont {Cui}}, \bibinfo {author}
  {\bibfnamefont {A.~S.}\ \bibnamefont {Bleich}}, \bibinfo {author}
  {\bibfnamefont {J.~G.}\ \bibnamefont {Analytis}}, \bibinfo {author}
  {\bibfnamefont {I.~R.}\ \bibnamefont {Fisher}}, \ and\ \bibinfo {author}
  {\bibfnamefont {D.}~\bibnamefont {Goldhaber-Gordon}},\ }\href {\doibase
  10.1103/PhysRevLett.109.056803} {\bibfield  {journal} {\bibinfo  {journal}
  {Phys. Rev. Lett.}\ }\textbf {\bibinfo {volume} {109}},\ \bibinfo {pages}
  {056803} (\bibinfo {year} {2012})}\BibitemShut {NoStop}%
\bibitem [{\citenamefont {Veldhorst}\ \emph {et~al.}(2012)\citenamefont
  {Veldhorst}, \citenamefont {Snelder}, \citenamefont {Hoek}, \citenamefont
  {Gang}, \citenamefont {Guduru}, \citenamefont {Wang}, \citenamefont
  {Zeitler}, \citenamefont {van~der Wiel}, \citenamefont {Golubov},
  \citenamefont {Hilgenkamp},\ and\ \citenamefont
  {Brinkman}}]{veldhorst2012josephson}%
  \BibitemOpen
  \bibfield  {author} {\bibinfo {author} {\bibfnamefont {M.}~\bibnamefont
  {Veldhorst}}, \bibinfo {author} {\bibfnamefont {M.}~\bibnamefont {Snelder}},
  \bibinfo {author} {\bibfnamefont {M.}~\bibnamefont {Hoek}}, \bibinfo {author}
  {\bibfnamefont {T.}~\bibnamefont {Gang}}, \bibinfo {author} {\bibfnamefont
  {V.~K.}\ \bibnamefont {Guduru}}, \bibinfo {author} {\bibfnamefont {X.~L.}\
  \bibnamefont {Wang}}, \bibinfo {author} {\bibfnamefont {U.}~\bibnamefont
  {Zeitler}}, \bibinfo {author} {\bibfnamefont {W.~G.}\ \bibnamefont {van~der
  Wiel}}, \bibinfo {author} {\bibfnamefont {A.~A.}\ \bibnamefont {Golubov}},
  \bibinfo {author} {\bibfnamefont {H.}~\bibnamefont {Hilgenkamp}}, \ and\
  \bibinfo {author} {\bibfnamefont {A.}~\bibnamefont {Brinkman}},\ }\href
  {\doibase 10.1038/nmat3255} {\bibfield  {journal} {\bibinfo  {journal} {Nat
  Mater}\ }\textbf {\bibinfo {volume} {11}},\ \bibinfo {pages} {417} (\bibinfo
  {year} {2012})}\BibitemShut {NoStop}%
\bibitem [{\citenamefont {Rokhinson}\ \emph {et~al.}(2012)\citenamefont
  {Rokhinson}, \citenamefont {Liu},\ and\ \citenamefont
  {Furdyna}}]{rokhinson2012thefractional}%
  \BibitemOpen
  \bibfield  {author} {\bibinfo {author} {\bibfnamefont {L.~P.}\ \bibnamefont
  {Rokhinson}}, \bibinfo {author} {\bibfnamefont {X.}~\bibnamefont {Liu}}, \
  and\ \bibinfo {author} {\bibfnamefont {J.~K.}\ \bibnamefont {Furdyna}},\
  }\href {\doibase 10.1038/nphys2429} {\bibfield  {journal} {\bibinfo
  {journal} {Nat Phys}\ }\textbf {\bibinfo {volume} {8}},\ \bibinfo {pages}
  {795} (\bibinfo {year} {2012})}\BibitemShut {NoStop}%
\bibitem [{\citenamefont {Seradjeh}\ \emph {et~al.}(2009)\citenamefont
  {Seradjeh}, \citenamefont {Moore},\ and\ \citenamefont
  {Franz}}]{seradjeh2009exciton}%
  \BibitemOpen
  \bibfield  {author} {\bibinfo {author} {\bibfnamefont {B.}~\bibnamefont
  {Seradjeh}}, \bibinfo {author} {\bibfnamefont {J.~E.}\ \bibnamefont {Moore}},
  \ and\ \bibinfo {author} {\bibfnamefont {M.}~\bibnamefont {Franz}},\ }\href
  {\doibase 10.1103/PhysRevLett.103.066402} {\bibfield  {journal} {\bibinfo
  {journal} {Phys. Rev. Lett.}\ }\textbf {\bibinfo {volume} {103}},\ \bibinfo
  {pages} {066402} (\bibinfo {year} {2009})}\BibitemShut {NoStop}%
\bibitem [{\citenamefont {Yu}\ \emph {et~al.}(2010)\citenamefont {Yu},
  \citenamefont {Zhang}, \citenamefont {Zhang}, \citenamefont {Zhang},
  \citenamefont {Dai},\ and\ \citenamefont {Fang}}]{yu2010quantized}%
  \BibitemOpen
  \bibfield  {author} {\bibinfo {author} {\bibfnamefont {R.}~\bibnamefont
  {Yu}}, \bibinfo {author} {\bibfnamefont {W.}~\bibnamefont {Zhang}}, \bibinfo
  {author} {\bibfnamefont {H.-J.}\ \bibnamefont {Zhang}}, \bibinfo {author}
  {\bibfnamefont {S.-C.}\ \bibnamefont {Zhang}}, \bibinfo {author}
  {\bibfnamefont {X.}~\bibnamefont {Dai}}, \ and\ \bibinfo {author}
  {\bibfnamefont {Z.}~\bibnamefont {Fang}},\ }\href {\doibase
  10.1126/science.1187485} {\bibfield  {journal} {\bibinfo  {journal}
  {Science}\ }\textbf {\bibinfo {volume} {329}},\ \bibinfo {pages} {61}
  (\bibinfo {year} {2010})},\ \bibinfo {note} {{PMID:} 20522741}\BibitemShut
  {NoStop}%
\bibitem [{\citenamefont {Chang}\ \emph {et~al.}(2013)\citenamefont {Chang},
  \citenamefont {Zhang}, \citenamefont {Feng}, \citenamefont {Shen},
  \citenamefont {Zhang}, \citenamefont {Guo}, \citenamefont {Li}, \citenamefont
  {Ou}, \citenamefont {Wei}, \citenamefont {Wang}, \citenamefont {Ji},
  \citenamefont {Feng}, \citenamefont {Ji}, \citenamefont {Chen}, \citenamefont
  {Jia}, \citenamefont {Dai}, \citenamefont {Fang}, \citenamefont {Zhang},
  \citenamefont {He}, \citenamefont {Wang}, \citenamefont {Lu}, \citenamefont
  {Ma},\ and\ \citenamefont {Xue}}]{chang2013experimental}%
  \BibitemOpen
  \bibfield  {author} {\bibinfo {author} {\bibfnamefont {C.-Z.}\ \bibnamefont
  {Chang}}, \bibinfo {author} {\bibfnamefont {J.}~\bibnamefont {Zhang}},
  \bibinfo {author} {\bibfnamefont {X.}~\bibnamefont {Feng}}, \bibinfo {author}
  {\bibfnamefont {J.}~\bibnamefont {Shen}}, \bibinfo {author} {\bibfnamefont
  {Z.}~\bibnamefont {Zhang}}, \bibinfo {author} {\bibfnamefont
  {M.}~\bibnamefont {Guo}}, \bibinfo {author} {\bibfnamefont {K.}~\bibnamefont
  {Li}}, \bibinfo {author} {\bibfnamefont {Y.}~\bibnamefont {Ou}}, \bibinfo
  {author} {\bibfnamefont {P.}~\bibnamefont {Wei}}, \bibinfo {author}
  {\bibfnamefont {L.-L.}\ \bibnamefont {Wang}}, \bibinfo {author}
  {\bibfnamefont {Z.-Q.}\ \bibnamefont {Ji}}, \bibinfo {author} {\bibfnamefont
  {Y.}~\bibnamefont {Feng}}, \bibinfo {author} {\bibfnamefont {S.}~\bibnamefont
  {Ji}}, \bibinfo {author} {\bibfnamefont {X.}~\bibnamefont {Chen}}, \bibinfo
  {author} {\bibfnamefont {J.}~\bibnamefont {Jia}}, \bibinfo {author}
  {\bibfnamefont {X.}~\bibnamefont {Dai}}, \bibinfo {author} {\bibfnamefont
  {Z.}~\bibnamefont {Fang}}, \bibinfo {author} {\bibfnamefont {S.-C.}\
  \bibnamefont {Zhang}}, \bibinfo {author} {\bibfnamefont {K.}~\bibnamefont
  {He}}, \bibinfo {author} {\bibfnamefont {Y.}~\bibnamefont {Wang}}, \bibinfo
  {author} {\bibfnamefont {L.}~\bibnamefont {Lu}}, \bibinfo {author}
  {\bibfnamefont {X.-C.}\ \bibnamefont {Ma}}, \ and\ \bibinfo {author}
  {\bibfnamefont {Q.-K.}\ \bibnamefont {Xue}},\ }\href {\doibase
  10.1126/science.1234414} {\bibfield  {journal} {\bibinfo  {journal}
  {Science}\ }\textbf {\bibinfo {volume} {340}},\ \bibinfo {pages} {167}
  (\bibinfo {year} {2013})}\BibitemShut {NoStop}%
\bibitem [{SMO()}]{SMO}%
  \BibitemOpen
  \href@noop {} {\enquote {\bibinfo {title} {See supplemental material at [url]
  for additional arpes data, data on a third device, hall effect data, gating
  behavior of conductance fluctuations, a detailed derivation of the charging
  model, and additional temperature- and magnetic field-dependence data.}}\
  }\BibitemShut {NoStop}%
\bibitem [{\citenamefont {Sobota}\ \emph {et~al.}(2012)\citenamefont {Sobota},
  \citenamefont {Yang}, \citenamefont {Analytis}, \citenamefont {Chen},
  \citenamefont {Fisher}, \citenamefont {Kirchmann},\ and\ \citenamefont
  {Shen}}]{sobota2012ultrafast}%
  \BibitemOpen
  \bibfield  {author} {\bibinfo {author} {\bibfnamefont {J.~A.}\ \bibnamefont
  {Sobota}}, \bibinfo {author} {\bibfnamefont {S.}~\bibnamefont {Yang}},
  \bibinfo {author} {\bibfnamefont {J.~G.}\ \bibnamefont {Analytis}}, \bibinfo
  {author} {\bibfnamefont {Y.~L.}\ \bibnamefont {Chen}}, \bibinfo {author}
  {\bibfnamefont {I.~R.}\ \bibnamefont {Fisher}}, \bibinfo {author}
  {\bibfnamefont {P.~S.}\ \bibnamefont {Kirchmann}}, \ and\ \bibinfo {author}
  {\bibfnamefont {Z.-X.}\ \bibnamefont {Shen}},\ }\href {\doibase
  10.1103/PhysRevLett.108.117403} {\bibfield  {journal} {\bibinfo  {journal}
  {Phys. Rev. Lett.}\ }\textbf {\bibinfo {volume} {108}},\ \bibinfo {pages}
  {117403} (\bibinfo {year} {2012})}\BibitemShut {NoStop}%
\bibitem [{\citenamefont {Wang}\ \emph {et~al.}(2012)\citenamefont {Wang},
  \citenamefont {Hsieh}, \citenamefont {Sie}, \citenamefont {Steinberg},
  \citenamefont {Gardner}, \citenamefont {Lee}, \citenamefont
  {Jarillo-Herrero},\ and\ \citenamefont {Gedik}}]{wang2012measurement}%
  \BibitemOpen
  \bibfield  {author} {\bibinfo {author} {\bibfnamefont {Y.~H.}\ \bibnamefont
  {Wang}}, \bibinfo {author} {\bibfnamefont {D.}~\bibnamefont {Hsieh}},
  \bibinfo {author} {\bibfnamefont {E.~J.}\ \bibnamefont {Sie}}, \bibinfo
  {author} {\bibfnamefont {H.}~\bibnamefont {Steinberg}}, \bibinfo {author}
  {\bibfnamefont {D.~R.}\ \bibnamefont {Gardner}}, \bibinfo {author}
  {\bibfnamefont {Y.~S.}\ \bibnamefont {Lee}}, \bibinfo {author} {\bibfnamefont
  {P.}~\bibnamefont {Jarillo-Herrero}}, \ and\ \bibinfo {author} {\bibfnamefont
  {N.}~\bibnamefont {Gedik}},\ }\href {\doibase 10.1103/PhysRevLett.109.127401}
  {\bibfield  {journal} {\bibinfo  {journal} {Phys. Rev. Lett.}\ }\textbf
  {\bibinfo {volume} {109}},\ \bibinfo {pages} {127401} (\bibinfo {year}
  {2012})}\BibitemShut {NoStop}%
\bibitem [{\citenamefont {Kim}\ \emph {et~al.}(2014)\citenamefont {Kim},
  \citenamefont {Yoshizawa}, \citenamefont {Ishida}, \citenamefont {Eto},
  \citenamefont {Segawa}, \citenamefont {Ando}, \citenamefont {Shin},\ and\
  \citenamefont {Komori}}]{kim2014robustprotection}%
  \BibitemOpen
  \bibfield  {author} {\bibinfo {author} {\bibfnamefont {S.}~\bibnamefont
  {Kim}}, \bibinfo {author} {\bibfnamefont {S.}~\bibnamefont {Yoshizawa}},
  \bibinfo {author} {\bibfnamefont {Y.}~\bibnamefont {Ishida}}, \bibinfo
  {author} {\bibfnamefont {K.}~\bibnamefont {Eto}}, \bibinfo {author}
  {\bibfnamefont {K.}~\bibnamefont {Segawa}}, \bibinfo {author} {\bibfnamefont
  {Y.}~\bibnamefont {Ando}}, \bibinfo {author} {\bibfnamefont {S.}~\bibnamefont
  {Shin}}, \ and\ \bibinfo {author} {\bibfnamefont {F.}~\bibnamefont
  {Komori}},\ }\href {\doibase 10.1103/PhysRevLett.112.136802} {\bibfield
  {journal} {\bibinfo  {journal} {Phys. Rev. Lett.}\ }\textbf {\bibinfo
  {volume} {112}},\ \bibinfo {pages} {136802} (\bibinfo {year}
  {2014})}\BibitemShut {NoStop}%
\bibitem [{\citenamefont {Dean}\ \emph {et~al.}(2010)\citenamefont {Dean},
  \citenamefont {Young}, \citenamefont {Meric}, \citenamefont {Lee},
  \citenamefont {Wang}, \citenamefont {Sorgenfrei}, \citenamefont {Watanabe},
  \citenamefont {Taniguchi}, \citenamefont {Kim}, \citenamefont {Shepard},\
  and\ \citenamefont {Hone}}]{dean2010boronnitride}%
  \BibitemOpen
  \bibfield  {author} {\bibinfo {author} {\bibfnamefont {C.~R.}\ \bibnamefont
  {Dean}}, \bibinfo {author} {\bibfnamefont {A.~F.}\ \bibnamefont {Young}},
  \bibinfo {author} {\bibfnamefont {I.}~\bibnamefont {Meric}}, \bibinfo
  {author} {\bibfnamefont {C.}~\bibnamefont {Lee}}, \bibinfo {author}
  {\bibfnamefont {L.}~\bibnamefont {Wang}}, \bibinfo {author} {\bibfnamefont
  {S.}~\bibnamefont {Sorgenfrei}}, \bibinfo {author} {\bibfnamefont
  {K.}~\bibnamefont {Watanabe}}, \bibinfo {author} {\bibfnamefont
  {T.}~\bibnamefont {Taniguchi}}, \bibinfo {author} {\bibfnamefont
  {P.}~\bibnamefont {Kim}}, \bibinfo {author} {\bibfnamefont {K.~L.}\
  \bibnamefont {Shepard}}, \ and\ \bibinfo {author} {\bibfnamefont
  {J.}~\bibnamefont {Hone}},\ }\href {\doibase 10.1038/nnano.2010.172}
  {\bibfield  {journal} {\bibinfo  {journal} {Nat Nano}\ }\textbf {\bibinfo
  {volume} {5}},\ \bibinfo {pages} {722} (\bibinfo {year} {2010})}\BibitemShut
  {NoStop}%
\bibitem [{\citenamefont {Adam}\ \emph {et~al.}(2012)\citenamefont {Adam},
  \citenamefont {Hwang},\ and\ \citenamefont
  {Das~Sarma}}]{adam2012twodimensional}%
  \BibitemOpen
  \bibfield  {author} {\bibinfo {author} {\bibfnamefont {S.}~\bibnamefont
  {Adam}}, \bibinfo {author} {\bibfnamefont {E.~H.}\ \bibnamefont {Hwang}}, \
  and\ \bibinfo {author} {\bibfnamefont {S.}~\bibnamefont {Das~Sarma}},\ }\href
  {\doibase 10.1103/PhysRevB.85.235413} {\bibfield  {journal} {\bibinfo
  {journal} {Phys. Rev. B}\ }\textbf {\bibinfo {volume} {85}},\ \bibinfo
  {pages} {235413} (\bibinfo {year} {2012})}\BibitemShut {NoStop}%
\bibitem [{\citenamefont {Xue}\ \emph {et~al.}(2011)\citenamefont {Xue},
  \citenamefont {Sanchez-Yamagishi}, \citenamefont {Bulmash}, \citenamefont
  {Jacquod}, \citenamefont {Deshpande}, \citenamefont {Watanabe}, \citenamefont
  {Taniguchi}, \citenamefont {Jarillo-Herrero},\ and\ \citenamefont
  {{LeRoy}}}]{xue2011scanning}%
  \BibitemOpen
  \bibfield  {author} {\bibinfo {author} {\bibfnamefont {J.}~\bibnamefont
  {Xue}}, \bibinfo {author} {\bibfnamefont {J.}~\bibnamefont
  {Sanchez-Yamagishi}}, \bibinfo {author} {\bibfnamefont {D.}~\bibnamefont
  {Bulmash}}, \bibinfo {author} {\bibfnamefont {P.}~\bibnamefont {Jacquod}},
  \bibinfo {author} {\bibfnamefont {A.}~\bibnamefont {Deshpande}}, \bibinfo
  {author} {\bibfnamefont {K.}~\bibnamefont {Watanabe}}, \bibinfo {author}
  {\bibfnamefont {T.}~\bibnamefont {Taniguchi}}, \bibinfo {author}
  {\bibfnamefont {P.}~\bibnamefont {Jarillo-Herrero}}, \ and\ \bibinfo {author}
  {\bibfnamefont {B.~J.}\ \bibnamefont {{LeRoy}}},\ }\href {\doibase
  10.1038/nmat2968} {\bibfield  {journal} {\bibinfo  {journal} {Nat Mater}\
  }\textbf {\bibinfo {volume} {10}},\ \bibinfo {pages} {282} (\bibinfo {year}
  {2011})}\BibitemShut {NoStop}%
\bibitem [{\citenamefont {Checkelsky}\ \emph {et~al.}(2012)\citenamefont
  {Checkelsky}, \citenamefont {Ye}, \citenamefont {Onose}, \citenamefont
  {Iwasa},\ and\ \citenamefont {Tokura}}]{checkelsky2012diracfermionmediated}%
  \BibitemOpen
  \bibfield  {author} {\bibinfo {author} {\bibfnamefont {J.~G.}\ \bibnamefont
  {Checkelsky}}, \bibinfo {author} {\bibfnamefont {J.}~\bibnamefont {Ye}},
  \bibinfo {author} {\bibfnamefont {Y.}~\bibnamefont {Onose}}, \bibinfo
  {author} {\bibfnamefont {Y.}~\bibnamefont {Iwasa}}, \ and\ \bibinfo {author}
  {\bibfnamefont {Y.}~\bibnamefont {Tokura}},\ }\href {\doibase
  10.1038/nphys2388} {\bibfield  {journal} {\bibinfo  {journal} {Nature
  Physics}\ }\textbf {\bibinfo {volume} {8}},\ \bibinfo {pages} {729} (\bibinfo
  {year} {2012})}\BibitemShut {NoStop}%
\bibitem [{\citenamefont {Kim}\ \emph {et~al.}(2012{\natexlab{b}})\citenamefont
  {Kim}, \citenamefont {Jo}, \citenamefont {Dillen}, \citenamefont {Ferrer},
  \citenamefont {Fallahazad}, \citenamefont {Yao}, \citenamefont {Banerjee},\
  and\ \citenamefont {Tutuc}}]{kim2012directmeasurement}%
  \BibitemOpen
  \bibfield  {author} {\bibinfo {author} {\bibfnamefont {S.}~\bibnamefont
  {Kim}}, \bibinfo {author} {\bibfnamefont {I.}~\bibnamefont {Jo}}, \bibinfo
  {author} {\bibfnamefont {D.~C.}\ \bibnamefont {Dillen}}, \bibinfo {author}
  {\bibfnamefont {D.~A.}\ \bibnamefont {Ferrer}}, \bibinfo {author}
  {\bibfnamefont {B.}~\bibnamefont {Fallahazad}}, \bibinfo {author}
  {\bibfnamefont {Z.}~\bibnamefont {Yao}}, \bibinfo {author} {\bibfnamefont
  {S.~K.}\ \bibnamefont {Banerjee}}, \ and\ \bibinfo {author} {\bibfnamefont
  {E.}~\bibnamefont {Tutuc}},\ }\href {\doibase 10.1103/PhysRevLett.108.116404}
  {\bibfield  {journal} {\bibinfo  {journal} {Phys. Rev. Lett.}\ }\textbf
  {\bibinfo {volume} {108}},\ \bibinfo {pages} {116404} (\bibinfo {year}
  {2012}{\natexlab{b}})}\BibitemShut {NoStop}%
\bibitem [{\citenamefont {{Collaboration: Authors and editors of the volumes
  {III/17E-17F-41C}}}()}]{collaboration:authorsantimony}%
  \BibitemOpen
  \bibfield  {author} {\bibinfo {author} {\bibnamefont {{Collaboration: Authors
  and editors of the volumes {III/17E-17F-41C}}}},\ }in\ \href
  {http://www.springermaterials.com/index/chapterdoi/10.1007/10681727_1045}
  {\emph {\bibinfo {booktitle} {Non-Tetrahedrally Bonded Elements and Binary
  Compounds I}}},\ Vol.\ \bibinfo {volume} {{41C}},\ \bibinfo {editor} {edited
  by\ \bibinfo {editor} {\bibfnamefont {O.}~\bibnamefont {Madelung}}, \bibinfo
  {editor} {\bibfnamefont {U.}~\bibnamefont {R{\"o}ssler}}, \ and\ \bibinfo
  {editor} {\bibfnamefont {M.}~\bibnamefont {Schulz}}}\ (\bibinfo  {publisher}
  {Springer-Verlag},\ \bibinfo {address} {{Berlin/Heidelberg}})\ pp.\ \bibinfo
  {pages} {1--4}\BibitemShut {NoStop}%
\bibitem [{\citenamefont {Petzelt}\ and\ \citenamefont
  {Grigas}(1973)}]{petzelt1973farinfrared}%
  \BibitemOpen
  \bibfield  {author} {\bibinfo {author} {\bibfnamefont {J.}~\bibnamefont
  {Petzelt}}\ and\ \bibinfo {author} {\bibfnamefont {J.}~\bibnamefont
  {Grigas}},\ }\href {\doibase 10.1080/00150197308235780} {\bibfield  {journal}
  {\bibinfo  {journal} {Ferroelectrics}\ }\textbf {\bibinfo {volume} {5}},\
  \bibinfo {pages} {59} (\bibinfo {year} {1973})}\BibitemShut {NoStop}%
\bibitem [{\citenamefont {Richter}\ and\ \citenamefont
  {Becker}(1977)}]{richter1977araman}%
  \BibitemOpen
  \bibfield  {author} {\bibinfo {author} {\bibfnamefont {W.}~\bibnamefont
  {Richter}}\ and\ \bibinfo {author} {\bibfnamefont {C.~R.}\ \bibnamefont
  {Becker}},\ }\href {\doibase 10.1002/pssb.2220840226} {\bibfield  {journal}
  {\bibinfo  {journal} {physica status solidi (b)}\ }\textbf {\bibinfo {volume}
  {84}},\ \bibinfo {pages} {619{\textendash}628} (\bibinfo {year}
  {1977})}\BibitemShut {NoStop}%
\bibitem [{\citenamefont {Skinner}\ \emph {et~al.}(2013)\citenamefont
  {Skinner}, \citenamefont {Chen},\ and\ \citenamefont
  {Shklovskii}}]{skinner2013effects}%
  \BibitemOpen
  \bibfield  {author} {\bibinfo {author} {\bibfnamefont {B.}~\bibnamefont
  {Skinner}}, \bibinfo {author} {\bibfnamefont {T.}~\bibnamefont {Chen}}, \
  and\ \bibinfo {author} {\bibfnamefont {B.~I.}\ \bibnamefont {Shklovskii}},\
  }\href {\doibase 10.1134/S1063776113110150} {\bibfield  {journal} {\bibinfo
  {journal} {J. Exp. The. Phys.}\ }\textbf {\bibinfo {volume} {117}},\ \bibinfo
  {pages} {579} (\bibinfo {year} {2013})}\BibitemShut {NoStop}%
\bibitem [{\citenamefont {Taskin}\ and\ \citenamefont
  {Ando}(2011)}]{taskin2011berryphase}%
  \BibitemOpen
  \bibfield  {author} {\bibinfo {author} {\bibfnamefont {A.~A.}\ \bibnamefont
  {Taskin}}\ and\ \bibinfo {author} {\bibfnamefont {Y.}~\bibnamefont {Ando}},\
  }\href {\doibase 10.1103/PhysRevB.84.035301} {\bibfield  {journal} {\bibinfo
  {journal} {Phys. Rev. B}\ }\textbf {\bibinfo {volume} {84}},\ \bibinfo
  {pages} {035301} (\bibinfo {year} {2011})}\BibitemShut {NoStop}%
\bibitem [{\citenamefont {Oostinga}\ \emph {et~al.}(2008)\citenamefont
  {Oostinga}, \citenamefont {Heersche}, \citenamefont {Liu}, \citenamefont
  {Morpurgo},\ and\ \citenamefont {Vandersypen}}]{oostinga2008gateinduced}%
  \BibitemOpen
  \bibfield  {author} {\bibinfo {author} {\bibfnamefont {J.~B.}\ \bibnamefont
  {Oostinga}}, \bibinfo {author} {\bibfnamefont {H.~B.}\ \bibnamefont
  {Heersche}}, \bibinfo {author} {\bibfnamefont {X.}~\bibnamefont {Liu}},
  \bibinfo {author} {\bibfnamefont {A.~F.}\ \bibnamefont {Morpurgo}}, \ and\
  \bibinfo {author} {\bibfnamefont {L.~M.~K.}\ \bibnamefont {Vandersypen}},\
  }\href {\doibase 10.1038/nmat2082} {\bibfield  {journal} {\bibinfo  {journal}
  {Nat Mater}\ }\textbf {\bibinfo {volume} {7}},\ \bibinfo {pages} {151}
  (\bibinfo {year} {2008})}\BibitemShut {NoStop}%
\bibitem [{\citenamefont {Sushkov}\ \emph {et~al.}(2010)\citenamefont
  {Sushkov}, \citenamefont {Jenkins}, \citenamefont {Schmadel}, \citenamefont
  {Butch}, \citenamefont {Paglione},\ and\ \citenamefont
  {Drew}}]{sushkov2010farinfrared}%
  \BibitemOpen
  \bibfield  {author} {\bibinfo {author} {\bibfnamefont {A.~B.}\ \bibnamefont
  {Sushkov}}, \bibinfo {author} {\bibfnamefont {G.~S.}\ \bibnamefont
  {Jenkins}}, \bibinfo {author} {\bibfnamefont {D.~C.}\ \bibnamefont
  {Schmadel}}, \bibinfo {author} {\bibfnamefont {N.~P.}\ \bibnamefont {Butch}},
  \bibinfo {author} {\bibfnamefont {J.}~\bibnamefont {Paglione}}, \ and\
  \bibinfo {author} {\bibfnamefont {H.~D.}\ \bibnamefont {Drew}},\ }\href
  {\doibase 10.1103/PhysRevB.82.125110} {\bibfield  {journal} {\bibinfo
  {journal} {Phys. Rev. B}\ }\textbf {\bibinfo {volume} {82}},\ \bibinfo
  {pages} {125110} (\bibinfo {year} {2010})}\BibitemShut {NoStop}%
\bibitem [{\citenamefont {{LaForge}}\ \emph {et~al.}(2010)\citenamefont
  {{LaForge}}, \citenamefont {Frenzel}, \citenamefont {Pursley}, \citenamefont
  {Lin}, \citenamefont {Liu}, \citenamefont {Shi},\ and\ \citenamefont
  {Basov}}]{laforge2010optical}%
  \BibitemOpen
  \bibfield  {author} {\bibinfo {author} {\bibfnamefont {A.~D.}\ \bibnamefont
  {{LaForge}}}, \bibinfo {author} {\bibfnamefont {A.}~\bibnamefont {Frenzel}},
  \bibinfo {author} {\bibfnamefont {B.~C.}\ \bibnamefont {Pursley}}, \bibinfo
  {author} {\bibfnamefont {T.}~\bibnamefont {Lin}}, \bibinfo {author}
  {\bibfnamefont {X.}~\bibnamefont {Liu}}, \bibinfo {author} {\bibfnamefont
  {J.}~\bibnamefont {Shi}}, \ and\ \bibinfo {author} {\bibfnamefont {D.~N.}\
  \bibnamefont {Basov}},\ }\href {\doibase 10.1103/PhysRevB.81.125120}
  {\bibfield  {journal} {\bibinfo  {journal} {Phys. Rev. B}\ }\textbf {\bibinfo
  {volume} {81}},\ \bibinfo {pages} {125120} (\bibinfo {year}
  {2010})}\BibitemShut {NoStop}%
\end{thebibliography}

\end{document}